\documentclass[review]{elsarticle}
\usepackage{lineno,hyperref}
\usepackage{comment,color,tikz,amssymb}
\usetikzlibrary{matrix, positioning, arrows}
\usepackage{amsfonts, amsthm, amsmath,mathpazo}
\usepgflibrary{shapes.arrows}
\usepackage{txfonts}
\usepackage{pxfonts}
\usepackage{tabularx}
\usepackage[hang,flushmargin]{footmisc}
\usetikzlibrary{matrix, positioning, arrows,decorations}
\usetikzlibrary{decorations.pathreplacing}
\tikzset{main node/.style={circle,fill=white,draw,minimum size=3cm,inner sep=1pt},}

\begin{document}

\begin{frontmatter}

\title{Reasons, Values, Stakeholders: A Philosophical Framework for Explainable Artificial Intelligence}

\author{Atoosa Kasirzadeh}
\address{atoosa.kasirzadeh@mail.utoronto.ca \\ University of Toronto (Toronto, Canada) \& Australian National University (Canberra, Australia)}
\date{}





\begin{abstract}
The societal and ethical implications of the use of opaque artificial intelligence systems for consequential decisions, such as welfare allocation and criminal justice, have generated a lively debate among multiple stakeholder groups, including computer scientists, ethicists, social scientists, policy makers, and end users. However, the lack of a common language or a multi-dimensional framework to appropriately bridge the technical, epistemic, and normative aspects of this debate prevents the discussion from being as productive as it could be. Drawing on the philosophical literature on the nature and value of explanations, this paper offers a multi-faceted framework that brings more conceptual precision to the present debate by (1) identifying the types of explanations that are most pertinent to artificial intelligence predictions, (2) recognizing the relevance and importance of social and ethical values for the evaluation of these explanations, and (3) demonstrating the importance of these explanations for incorporating a diversified approach to improving the design of truthful algorithmic ecosystems. The proposed philosophical framework thus lays the groundwork for establishing a pertinent connection between the technical and ethical aspects of artificial intelligence systems.
\end{abstract}


\end{frontmatter}


\section{Preliminaries}

Governments and private actors are seeking to leverage recent developments in artificial intelligence (AI) and machine learning (ML) algorithms for use in high-stakes decision making. These algorithms have begun to greatly impact human lives by guiding several consequential decision-making procedures such as hiring employees, assigning loans and credit scores, making medical diagnoses, and dealing with recidivism in the criminal justice system (see, for example, \citep{cabitza2017unintended,angwin2016machine,bodo2017tackling,fox2007argumentation,ribeiro2016should,veale2018clarity}). This enterprise is motivated by the idea that high-stakes decision problems are instances of learnable algorithmic tasks, such as pattern recognition, classification, and clustering.

ML algorithms have had some success in making accurate predictions of unobserved data in some domains, but many such systems remain opaque. This means that it is difficult for humans to have sufficient reasons to understand why an algorithmic outcome or another is obtained in a given circumstance. A possible silver bullet to alleviate this opacity would be to require the algorithms to explain themselves. On this basis, several academic and industrial research programs have been put into operation to develop the so-called ``explainable AI'' systems: AI systems that are able to explain themselves (see, for example, \cite{gunning2017explainable,adadi2018peeking}).\footnote{In this paper, I am primarily concerned with opaque problems of ML systems, as a particular kind of opaque AI prediction systems. The problem of opacity can also arise for other AI paradigms (see, for example, \cite{swartout1993explanation}). However, I use the term AI broadly, following the tendency in the research associated with the explainable AI paradigm.} More concretely, the development of explainable AI systems is motivated by several social and ethical reasons (see, for example, \cite{binns2018s,chen2014situation,chouldechova2017fairer,hayes2017improving,kemper2019transparent,kim2015interactive,lepri2018fair,mercado2016intelligent,hoffman2018explaining,zhang2020effect}): increasing societal acceptance of prediction-based decisions, establishing trust in the results of these decisions, making algorithms accountable to the public, removing the sources of algorithmic discrimination and unfairness, legitimizing the incorporation of AI algorithms in several decision contexts, and facilitating a fruitful conversation among stakeholders on the justification of the use of these algorithms in decision making. These considerations, in addition to their epistemic importance, point to a normative and social significance for AI explanations. An appropriate accommodation of these considerations requires finding a clear and comprehensive answer to the question of what AI explanations are, who their consumers are, and how they matter. 

Computer scientists have identified no well-formed definition of AI explainability as an all-purpose, monolithic concept \citep{lipton2016mythos,doshi2017towards}. However, they have developed several technical methods for making the AI systems explainable to some degree (for two recent surveys, see \citep{adadi2018peeking,guidotti2018survey}). Nevertheless, it is necessary to ground these technical methods in a conceptually precise analysis of the nature and value of explanations if the quality of claims about the ability of intelligent agents can be reconciled and assessed in explanations that incorporate social and normative significance. A framework for AI explanations that specify why and how they matter remains to be established. One promising approach to achieving this is seeking inspiration from the social sciences and philosophy. 

To achieve this, I begin by briefly reviewing the literature that describes the main efforts that have been made to identify the conceptual foundations of and requirements for explainability.

In Miller \cite{miller2019explanation}, the most elaborate survey to-date, primarily the causal aspects of everyday, local explanations for building explainable AI systems are examined. This survey, however, does not include any discussion of non-causal aspects of AI explanation.\footnote{For instance, \cite{miller2019explanation} (p.6) says: \vspace{-0.4cm} \begin{quote}
But what constitutes an explanation? This question has created a lot of debate in philosophy, but accounts of explanation both philosophical and psychology [sic] stress the importance of causality in explanation — that is, an explanation refers to causes [...]. There are, however, definitions of non-causal explanation [...]. These definitions [are] out of scope in this paper, and they present a different set of challenges to explainable AI.
\end{quote}} Zerilli et al. \cite{zerilli2019transparency} argue that explanations we demand from opaque algorithmic systems must be given from an intentional stance. A popular framework for the analysis of to cognitive systems is used by Zednik \cite{zednik2019solving} to map different analytic techniques (such as input heatmapping), as developed by the explainable AI researchers, to respond to the epistemic demands of six kinds of stakeholders (as suggested by Tomsett et al. \citep{tomsett2018interpretable}): developers, examiners, operators, executors, data subjects, and decision subjects. 

As such, this paper extends these works in pursuit of the following goal: although epistemic demands and everyday, local explanations are necessary to build an intelligent explainable system, they are not sufficient to ensure that it is built. This is because AI systems are value-laden across several dimensions. This means that ineliminable contextual value judgments must be made about AI algorithms that necessarily inform consequential decisions. Here, I show why and how these values matter for the evaluation of the acceptability and quality of AI explanations. Because inherently complex and complicated AI ecosystems are connected with various stakeholders, institutions, values, and norms, the simplicity and locality of an explanation should not be the only virtue sought in the design of explainable AI. Hence, the goal here is to offer a sufficiently rich framework for explainable AI that can be abridged and adjusted in relation to the risks and consequences attributed to different prediction contexts.

In other words, merely a philosophical framework (such as that of Zednik's \citep{zednik2019solving}) for the correspondence between analytical methods of explainable AI and the epistemic demands of certain stakeholders cannot provide sufficient resources for the critical engagement of those who receive explanations from AI systems. More specifically, such a framework has two limitations. First, I argue that, in addition to epistemic demands, the social and normative values matter to developing a philosophical framework for explainable AI systems. Thus, I argue for a broadened scope of the explanations of interest for the design of an explainable system. Second, a simple mapping between analytical methods for explainable AI and the six stakeholder types excludes many significant stakeholders (including ethicists, policy-makers, and social and political scientists) who also assess AI explanations in relation to normative and social significance. To overcome these limitations, I offer a philosophical framework for building explainable AI systems that has sufficient resources to enable explicit articulation of why and how social, ethical, and epistemic values matter in terms of asserting the relevance of a range of explanation types and stakeholders in explaining AI outputs.

This paper broadens the scope and relevance of current frameworks for AI explanations to accommodate the significance of explainable AI and its role in ensuring democratic and ethical algorithmic decisions. In such a framework, the usefulness of AI explanations must be evaluated in relation to a variety of epistemic, normative, and social demands. These fall along the three axes Reasons, Values, and Stakeholders, in what I call the RVS framework. It includes different levels of explanation. Each level makes sense of algorithmic outcomes in relation to different sets of reasons, values, and stakeholders. The novelty here is that the levels of explanation do not simply map onto the analytic methods as explanations or to the six stakeholder categories. Explanations go beyond analytic methods and relate to practical, ethical, and political reasoning and values which are required if explanations are to be made sense of by broader sets of stakeholders. This framework enables diversification and the inclusion of different epistemic and practical standpoints in AI ecosystems.

The remainder of this paper is structured as follows. In Section 2, I argue that, in addition to data-analytical explanations, design explanations that incorporate representational, mathematical, and optimality information are pertinent to the understanding of the predictions of AI systems, and therefore should be included in the design of explainable AI systems. In Section 3, I propose a philosophically-informed schema in which a variety of design and data-analytical information can find their place in explanations of AI outcomes. In Section 4, I present our philosophical framework for explainable AI by incorporating ethical and social values into the evaluation of the quality and relevance of AI explanations. The paper is concluded in Section 5 with some directions for future work.

\section{Kinds of explanation}

Let us consider an algorithm that assesses job candidates to recommend one as a future employee for company X. Nora, a competent candidate, applies for the job. Her application is rejected. Nora wants to know why she is rejected (Figure 1). She does not seek a just-so story that makes sense by organizing events into an intelligible whole. She wants the explanation to have a factual foundation and to be empirically or mathematically true.

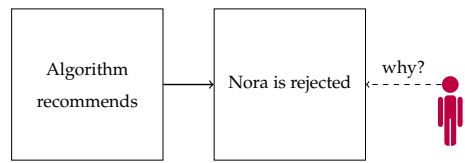
\begin{figure}[h]
\centering
\resizebox{6cm}{!}{\begin{tikzpicture}[every text node part/.style={align=center}]
    \node[main node, rectangle, inner sep=0.5ex] (1) {Algorithm \\ recommends};
    \node[main node, rectangle, inner sep=0.5ex] (5) [right = 1cm of 1] {Nora is rejected};
    \node[circle,fill=purple,minimum size=1mm] (head) [right=1.5cm of 5]{};
    \node[rounded corners=2pt,minimum height=1cm,minimum width=0.3mm,fill=purple,below = 1pt of head] (body) {};
    \draw[line width=1mm,purple,round cap-round cap] ([shift={(2pt,-1pt)}]body.north east) --++(-90:6mm);
    \draw[line width=1mm,purple,round cap-round cap] ([shift={(-2pt,-1pt)}]body.north west)--++(-90:6mm);
    \draw[thick,white,-round cap] (body.south) --++(90:5.5mm);
    \path[->,thick, below=8pt]
    (1) edge node {} (5);
    \path[->,dashed, right=8pt]
    (head) edge node [midway, above, sloped]{why?} (5); 
\end{tikzpicture}}

\caption{Algorithmic recommendation and the why-question.}
\end{figure}

This decision-making problem is exemplary of the social problems that might be resolved with a classification algorithm. Figure 1 can be expanded to Figure 2 to show a more fine-grained characterization of the decision process in terms of five main stages.

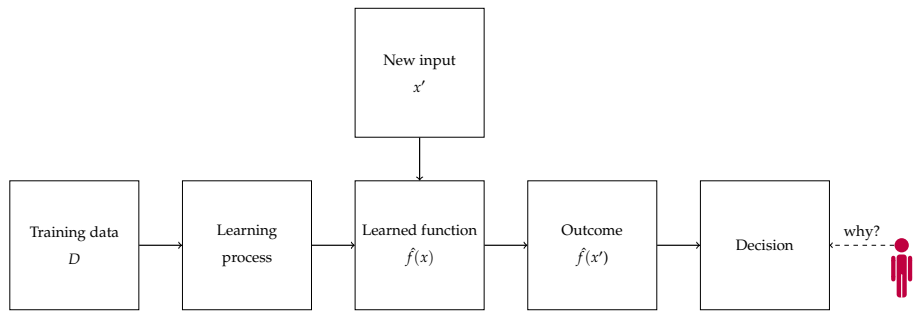
\begin{figure}[h]
\centering
\resizebox{12cm}{!}{\begin{tikzpicture}[every text node part/.style={align=center}]
    \node[main node, rectangle, inner sep=0.5ex] (1) {Training data \\ $D$};
    \node[main node, rectangle, inner sep=0.5ex] (2) [right = 1cm of 1] {Learning \\ process};
    \node[main node, rectangle, inner sep=0.5ex] (3) [right = 1cm of 2]  {Learned function \\ $\hat{f}(x)$};
    \node[main node, rectangle, inner sep=0.5ex] (4) [right = 1cm of 3] {Outcome \\ $\hat{f}(x^{\prime}$)};
    \node[main node, rectangle, inner sep=0.5ex] (5) [above = 1cm of 3]  {New input \\ $x^{\prime}$};
    \node[main node, rectangle, inner sep=0.5ex] (6) [right = 1cm of 4]  {Decision};
    \node[circle,fill=purple,minimum size=1mm] (head) [right=1.5cm of 6]{};
    \node[rounded corners=2pt,minimum height=1cm,minimum width=0.3mm,fill=purple,below = 1pt of head] (body) {};
    \draw[line width=1mm,purple,round cap-round cap] ([shift={(2pt,-1pt)}]body.north east) --++(-90:6mm);
    \draw[line width=1mm,purple,round cap-round cap] ([shift={(-2pt,-1pt)}]body.north west)--++(-90:6mm);
    \draw[thick,white,-round cap] (body.south) --++(90:5.5mm);
    \path[->,thick, above=8pt]
    (1) edge node {} (2);
    \path[->,thick, right=8pt]
    (2) edge node {} (3);  
     \path[->,thick, below=8pt]
    (3) edge node {} (4);
    \path[->,dashed, right=8pt]
    (head) edge node [midway, above, sloped]{why?} (6);
     \path[->,thick, right=8pt]
    (5) edge node {} (3);
     \path[->,thick, right=8pt]
    (4) edge node {} (6);
\end{tikzpicture}}
\caption{Prediction-based decision procedure and the why-question.}
\end{figure}

To make the matters more concrete, let us consider a deep supervised learning algorithm that makes hiring predictions (Figure 3).\footnote{We do not know anything of the detailed operations of many commercial ML systems due to intellectual property rights and the need to maintain trade secrets. As a result, we do not know the detailed operations of these algorithms in spite of their social impact. I acknowledge that the prevalent ML algorithms learn with different models, including deep neural networks, support-vector machines, and ensemble methods. For ease of exposition and without any loss of generality, this paper is primarily limited to discussion of examples of deep neural networks, which is a widely discussed type of truly opaque ML model.} That is, given a training data set composed of several labeled feature vectors, the algorithm learns to map applicants' CV features to their expected job performance. Given Nora's CV, the learned model predicts whether she will perform well if given the job. What explains the prediction that leads to Nora's rejection?

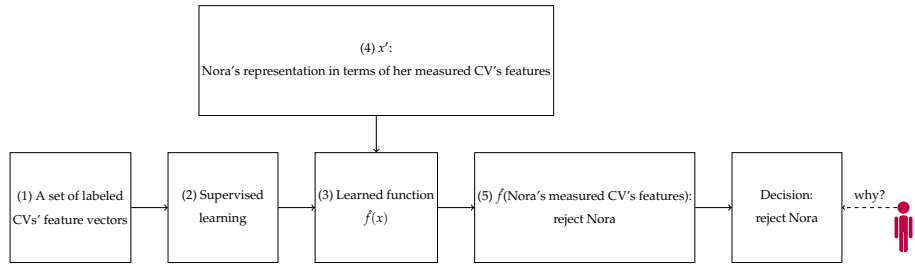
\begin{figure}[h]
\centering
\resizebox{12cm}{!}{\begin{tikzpicture}[every text node part/.style={align=center}]
    \node[main node, rectangle, inner sep=0.5ex] (1) {(1) A set of labeled \\ CVs' feature vectors};
    \node[main node, rectangle, inner sep=0.5ex] (2) [right = 1cm of 1] {(2) Supervised \\ learning};
    \node[main node, rectangle, inner sep=0.5ex] (3) [right = 1cm of 2]  {(3) Learned function \\ $\hat{f}(x)$};
        \node[main node, rectangle, inner sep=0.5ex] (4) [right = 1cm of 3] {(5) $\hat{f}$(Nora's measured CV's features): \\ reject Nora};
    \node[main node, rectangle, inner sep=0.5ex] (5) [above = 1cm of 3]  {(4) $x^\prime$: \\ Nora's representation in terms of her measured CV's features};
    \node[main node, rectangle, inner sep=0.5ex] (6) [right = 1cm of 4]  {Decision: \\ reject Nora};
    \node[circle,fill=purple,minimum size=1mm] (head) [right=1.5cm of 6]{};
    \node[rounded corners=2pt,minimum height=1cm,minimum width=0.3mm,fill=purple,below = 1pt of head] (body) {};
    \draw[line width=1mm,purple,round cap-round cap] ([shift={(2pt,-1pt)}]body.north east) --++(-90:6mm);
    \draw[line width=1mm,purple,round cap-round cap] ([shift={(-2pt,-1pt)}]body.north west)--++(-90:6mm);
    \draw[thick,white,-round cap] (body.south) --++(90:5.5mm);
    \path[->,thick, above=8pt]
    (1) edge node {} (2);
    \path[->,thick, right=8pt]
    (2) edge node {} (3);  
     \path[->,thick, below=8pt]
    (3) edge node {} (4);
    \path[->,dashed, right=8pt]
    (head) edge node [midway, above, sloped]{why?} (6); 
     \path[->,thick, right=8pt]
    (5) edge node {} (3);
     \path[->,thick, right=8pt]
    (4) edge node {} (6);
\end{tikzpicture}}
\caption{Prediction-based hiring decision.}
\end{figure}

For now, we are not considering ourselves to be bound by the epistemic needs of particular stakeholders. Rather, we take a broad and maximalist perspective, and we seek a multi-dimensional explanation for this algorithmic prediction. I tackle the attribution of stakeholders and the relevance of each kind of explanation for different normative and social considerations in the next section.

One natural way to cash out the explanation for why Nora is rejected is to decompose the explanatory information in terms of each of the five ML-relevant stages sketched in Figure 3. Collectively, the information in steps (1) -- (5) produces the determination of the prediction, and therefore explains how a particular algorithmic recommendation is obtained for Nora. 

Consider some potential responses to the five clusters of questions specified by (1) -- (5) in Figure 3. (1) What role does the size and inclusiveness of the training data play in recommending for Nora's rejection? What would have happened if the training data were different, incorporated different criteria for predicting applicant success, or were pre-processed with different standards? (2) How does the particular learning process of the system determine the prediction outcome? What would the algorithm's outcome be if the learning process had been different, if for example the algorithm had learned in an unsupervised fashion? (3) How does the algorithm's specific thinking style (i.e., way of reasoning) in relation to its particular constituents (loss function or hyper parameters, for example) impact what it predicts? (4) Why and how does this representation of Nora, in terms of quantified and measured input features, produce this recommendation? After all, it would seem perfectly reasonable to imagine that a different choice for the relevant input features would give a different prediction outcome. (5) Which of Nora's features cause, correlate with, or counterfactually depend upon the prediction for her rejection?\footnote{Several definitions of causation and counterfactual dependence (see, for example, \cite{sep-causation-counterfactual}) have been developed. However, they are outside the scope of this paper, and I do not examine them here.} 

The majority of work on the design of explainable AI systems hitherto has focused on methods that emphasize responses to explanatory questions (3) or (5), mainly by producing feature-based token explanations (i.e., explanations for particular cases, such as Nora's rejection in response to particular input features). There has also been some related work that designs explainability methods to approximate the logic of opaque reasoning in a deep neural network, such as by extracting a decision tree, or by visualizing the network's hidden layer activity. There approaches relate to a class of design-based AI explanations, that approximate the design of the system, as I discuss in Section 2.2.

These methods explain algorithmic prediction in terms of dependency patterns, namely, causal, counterfactual, correlational, and contrastive ones, relating salient input features and the algorithmic outcome. Let us review these approaches briefly before turning to the role of design explanations for algorithmic predictions.

\subsection{Feature-based token explanations}

Attempts at making AI explainable in terms of technical methods to extract dependency relations between the input and target features fall roughly into three categories (i) -- (iii).

(i) \emph{Correlational explanations} provide information on associations between certain specifically important input features and output predictions. These associations can be captured by using methods such as feature importance, a saliency map, a partial dependence plot, individual conditional expectation, or rule extraction to indicate how important the association of a given input feature (e.g., Nora's longevity at her workplace or the temperature) is to an output prediction (e.g., of Nora's duration of stay at the given job or the number of bikers in the streets) \cite{erhan2009visualizing,berk2013statistical,goldstein2015peeking,letham2015interpretable,ribeiro2016should,ribeiro2016model,sundararajan2017axiomatic,dabkowski2017real,lundberg2017unified,kim2018}. 

(ii) \emph{Causal and counterfactual explanations} reveal causal dependencies among input features and predicted outcomes. Causal explanations can be obtained with methods such as causal modeling or approximate decision trees \citep{hara2016making,wachter2017counterfactual,lakkaraju2017learning,zhao2019causal}. For instance, a causal explanation for why Nora was recommended for rejection might be that some of her input features, such as her ethnicity, led to a lower score for her predicted performance. 

(iii) \emph{Example-based explanations}, such as prototype explanations, generate information by comparing one data instance to a similar one or to a particular prototype \citep{kim2014bayesian,kim2016examples,li2018deep}. For an instance of this kind of explanation, consider Aron, who with a very similar CV to Nora but a longer job longevity, was given the job.

Thus, explanations of the kind (i) -- (iii) produce information on relevant and important input features for prediction outputs in particular scenarios of interest (Nora's rejection, for example). This information gives feature-based responses to explanatory questions (3) and (5). However, mere responses to these questions do not satisfy the need for multiple normative and social considerations (e.g., those that would allow trust in algorithmic predictions) that I outlined in the previous section, and that motivate the attempt to recognize what explainable AI is in the first place. For example, a feature-based explanation generated by a saliency map might observe how the algorithm might have reasoned in a particular case (such as in Nora's rejection) by highlighting the patterns that the model is looking at, but it does not explain (or try to explain) why we should trust this prediction. 

In the remainder of this section, I discuss three other kinds of explanation, that unlike feature-based token explanations, provide design-relevant information to respond how a certain prediction may be obtained. Consequently, these three explanatory types can respond to questions (1), (2), and (4).

\subsection{Design explanations}

ML algorithms incorporate an amalgamation of inductive reasoning and mathematical optimization to learn and perform a task, such as predicting the likelihood of the occurrence of an outcome. The particulars of the algorithmic design, such as the choice of loss function (e.g., least-square or exponential) and the learning style (e.g., supervised or unsupervised), determine the inference outputs of the algorithm. 

This section indicates that because the reasoning of an ML algorithm is \emph{defined} in terms of mathematical and optimality facts, these must govern and determine how the data are processed. Therefore, such facts must (partially) explain why a given algorithmic output is obtained. Consequently, the mathematical design of an algorithm provides mathematical explanations for why an algorithmic outcome can be achieved. A recent study, for example, shows how two different design approaches to ML algorithms (convolutional neural networks \citep{krizhevsky2012imagenet} and capsule networks \citep{sabour2017dynamic,kosiorek2019stacked}) differently explain why an image is classified one way rather than another, and in some cases, choosing a different mathematical design for an algorithm produces a significant difference to the generated output.

Decisions made on the basis of ML predictions encode the normative claim that algorithmic inference via certain mathematical and optimality constituents is the right way to produce outcome predictions because they play a central role in explaining why a given outcome for a deployed ML algorithm is obtained.

To make the significance of a mathematical explanation of explainable AI more concrete, I provide a simple example that may fix the relevant intuitions regarding the explanatory power of mathematical facts in explaining pricing in the insurance industry. 

\subsubsection{Mathematical and optimality explanations}

In the insurance industry, statistical methods are used to calculate risk premiums based on the past behavior of a certain population rather than of a single individual. The cost of one's insurance policy depends on one's assessed risk. This risk represents the likelihood of making an insurance claim, and the lower that risk, the lower the assigned premium. Insurers rely heavily on statistical rules governing large numbers and on the central limit theorem to determine how much they are likely to pay out by calculating risk premiums. For instance, the central limit theorem warrants the use of a Gaussian distribution for estimation and likelihood definition when sums of independent and identically distributed random variables are considered. The law of large numbers and the central limit theorem, as established mathematical facts, can be used to explain the pooling of losses as an insurance mechanism, and they also explain why larger numbers of policyholders strengthens the sustainability of an insurance company's business by reducing the probability that its pool will fail \citep{smith1994law}.

In a similar vein, consider a learning algorithm that classifies whether a tumor is malignant or benign. The explanation for the classification of a given tumor as malignant rather than benign is determined by calculating highly probable similarities to an analyzed statistical sample that suggest that the tumor is like other malignant tumors. Such probabilistic similarities are warranted through mathematical facts that are constitutive of their learning style from data sets. In sum, because such mathematical facts partially govern and influence the design of the decision procedure, they are among the explanatory factors indicating why one or another outcome is achieved. Returning to Nora's rejection, the recommendation is obtained from warranted statistical paths, in terms of which the ML algorithm is defined, designed, and governed. If a different set of mathematical assumptions is chosen to guide the algorithm's inferences, a different decision outcome would have been produced. Having established the explanatory relevance of certain mathematical facts for AI reasoning, I now make a case for the explanatory role of optimality results to the predictions of ML algorithms.

Most ML algorithms are defined to optimize a particular objective function (primarily a loss function) for the purposes of their training. Supervised learning algorithms, for instance, require an optimization problem to be resolved such that the total loss function for expected accuracy is minimized using methods such as the stochastic gradient descent. The loss function is minimized by continuously updating the weights in an artificial neural network, for example, using methods such as backpropagation.\footnote{The total loss of the network is the sum of loss for all output-layer nodes or the composition of all loss functions at the nodes of the output layer.} 

Backpropagation is based on four fundamental mathematical equations \citep{lecun2015deep}. Why should we trust that these equations will perform the task we want them to do? Simply, we can mathematically prove that the equations are true and successfully enable the algorithm to learn with a high degree of accuracy. Hence, mathematical proofs guarantee the efficiency of the use of backpropagation in an ML prediction problem. This shows that the optimality facts that warrant the algorithm's output also partially explain why that decision output is obtained. For instance, in case of Nora's hiring, the algorithm is designed to optimize certain success criteria, such as low probability of quitting a job for a (set of) chosen target variable(s). The choice of a particular loss function partially determines why Nora is rejected, so it (partially) explains this decision.

Indeed, adopting different learning algorithms requires making several pragmatic choices that guarantee the algorithmic performance and handling of computational complexity (e.g., how the algorithm's reasoning style should adjust weights, what the activation function should be at each node, and how many nodes and layers are required to set up the algorithm). If a different mathematical instantiation had been used (such as a different loss function or regularization), a different algorithmic output could be produced.

So far, I have discussed that in addition to the three kinds of feature-based token explanations -- correlational, causal, and example-based -- there are also design explanations that are pertinent to understanding AI outputs, and hence to the design of an explainable AI. The point here is that because these mathematical and optimality constituents define the algorithmic reasoning style, they impact how the outcome is determined. Hence, the mathematical constituents of the algorithm's inference-making must manifest themselves in the explanation that indicate why a particular algorithmic prediction is obtained. The philosophical literature on mathematical (and non-causal) explanations can help us better to ground the varieties of AI explanations within a unifying whole.

\subsubsection{Mathematical explanations: conceptual and philosophical foundations}

The question of what an explanation is has been central to philosophy for centuries. Contemporary philosophers have given extensive attention and care to characterizing the various aspects of explanation, in particular since Hempel \cite{hempel1948studies}. These discussions tend to focus on answering one of the following three questions. (1) Can explanations be reduced to causal explanations, or are there genuine cases of non-causal (e.g., mathematical) explanations? (2) Can we, and if so how can we, specify the necessary and sufficient conditions for an explanation? (3) How do non-causal explanations work, if they are at all valid?

In \cite{hempel1948studies, hempel1965aspects} an account of explanation is offered that emphasizes its argumentative nature. That account describes explanations as having two constituent parts: the thing to be explained, or the explanandum, and the thing that explains, or the explanans. It is stipulated \cite{hempel1948studies, hempel1965aspects} that the explanandum must logically follow from a set of explanans that contains at least one law-like generalization. The argumentative nature of the explanations generated by AI systems has been described by \citep{miller2019explanation, mittelstadt2019explaining}, among others. The account of explanation put forth by \cite{hempel1948studies, hempel1965aspects}, however, was ultimately found to be too permissive, allowing for irrelevant generalizations to be counted as instances of explanans.

Recently, several philosophers have argued that an important but hitherto unnoted element in the account of explanation is the provision of causal information. Thus, perhaps explanations are arguments that provide information about causal relations in the world \citep{salmon1984scientific, strevens2008depth}. If so, what role does mathematics play in explanations? Some have suggested that mathematics merely represents causal relations. However, not everyone agrees with this. Several philosophers relying on studies in the physical and social sciences, have argued that the discussion of what explanations are and what sort of information they likely provide also requires an unveiling of non-causal (e.g., mathematical) information as an explanans \citep{potochnik2007optimality,bokulich2011scientific,batterman2014minimal,chirimuuta2017explanation,lange2016because,pincock2007role,reutlinger2018explanation,rice2015moving,batterman2001devil}. Two simple examples indicate the insight that guides this perspective. 

To explain why a mother without the ability to break stones and with the desire to divide 23 stones among her three children cannot do so, one can appeal to the mathematical fact that 23 is not evenly divisible by 3. This mathematical fact, not any causal relation, is responsible for the mother's failure. For another example, consider the problem of the Seven Bridges of K{\"o}nigsberg \citep{pincock2007role}. This problem describes four parts of a town that are connected to each other by seven bridges. To explain why no one can cross all bridges exactly once on one walk, it is needed to appeal to a set of mathematical facts that were proven by Euler. First, an Euler path is defined as a path through a graph that crosses each arc exactly once. Then, it is proven that a graph has an Euler path only if either of the following two conditions is satisfied: the number of arcs that pass from each node is even, or exactly two nodes feature an odd total number of arcs passing from them. As neither of the conditions are satisfied in the case of the K{\"o}nigsberg problem, no Euler path can be found. This demonstration explains why the problem of crossing all seven bridges exactly once is insoluble. 

More broadly, some philosophers have argued that mathematical facts explain a variety of empirical phenomenon better than causal information does, across a range of sciences, from condensed matter physics to evolutionary and developmental biology or computational neuroscience \citep{potochnik2007optimality,bokulich2011scientific,batterman2014minimal,rice2015moving,lange2016because,chirimuuta2017explanation,reutlinger2018explanation}. Other examples include cases of optimality or statistical explanations where reference to an optimality or statistical fact is responsible for the explanation of empirical phenomena such as natural selection or genetic drift \citep{potochnik2007optimality,rice2015moving,lange2013really}. These philosophical discussions may place the nature of AI explanations regarding algorithmic decisions in the proper light. Below, I present two apparent examples of algorithmic outputs whose explanations must include mathematical information.

A Boeing 737 Max 8 aircraft crashed during Lion Air Flight 610, killing all 189 occupants \citep{johnston2019boeing}. This aircraft had an algorithmically controlled stability system for adjusting the angle of the airplane. A fault in the design of this algorithmic system has been reported to be among the main reasons why the crash happened. The design relied heavily on one sensor. When this sensor malfunctioned, an inaccurate signal was sent about the airplane's angle. To compensate, the algorithmic system pushed the airplane's nose down, but without sensor information indicating the success of the motor, the system continued to push the nose down, causing the crash.

What explanation do we give for the crash? Poor algorithmic design is among several explanations. If an algorithm with a more robust design had been used, a different decision outcome would have been obtained. Alternatively, consider the case of a linear programming optimization algorithm assigning a hospital's nurse scheduling plans. The reason that a particular set of nurse schedules is created is partly due to determining the results through the linear programming optimization. The point made concerning these two examples of algorithmic decision systems can be generalized. Because mathematical-optimization constituents are situated at the heart of any ML system, the decision outcome is governed and determined by what a set of mathematical bits mandate. 

We have argued that, in addition to feature-based token explanations, mathematical and optimality facts can also produce significant explanations of an AI system's output. In the remainder of this section, I briefly discuss the importance of two other design explanations, representational explanations and training data explanations, for the outcomes of an algorithmic system.

\subsection{Representational explanations}

Any ML system requires commitment, even if it is not explicitly acknowledged, to a choice of what features of the world will be taken as relevant for its learning. These features will be among the main building blocks for which the outcomes of the system of interest are analyzed and algorithmic predictions are made and explained. Therefore, the choice of features representing a system and their corresponding quantitative values also play a role in explaining why an algorithmic prediction is made: I call these the representational explanations.

Providing an appropriate representation of a target phenomenon in quantitatively measurable terms requires having clear and distinct ideas about the constitutive parts of the target whole and the relationships among the parts. This requires choosing what features of the target will be considered important and relevant to the representation. It also requires having access to appropriate quantitative measures of these inputs. These quantification demands are not the only objective way in which consequential scenarios can be tackled. Indeed, they are grounded in various economic motivations, and are rooted in particular perceptions of humans, values, and status \cite{mau2019metric}.

Such quantifications are a necessary building block for learning and performance in AI systems. Deep learning algorithms, for example, impose a special representational demand: a correspondence setup, in quantitative terms, between the input layer of the network and the source of the input layer data coming from the real world. The use of predictive deep neural networks requires all of the significant elements of the feature space (e.g., Nora's features) that are relevant to the decision-making situation to be reduced to a set of measured quantities.

Abstracting away from any real world entity (such as Nora's reality as a candidate) toward a particular set of representative features (Nora's longevity in previous employments or her ethnicity) that should be quantified requires commitment to a particular set of epistemological and ontological assumptions, as I discuss in Section 4. The representational choices made with respect to the decision problem constitute a response to explanatory question (4), Why and how does this representation of Nora, in terms of quantified and measured input features, produce this recommendation? The answer indicates the following assumption: the precondition for using these algorithms as decision-making tools is that the relevant features of Nora, as a person, must be reduced to quantifiable variables, measured, and then fed as input data to the algorithm. 

Why should we care about representational explanations? They provide responses to questions such as how to represent and why and how to measure pertinent features. These explanations bridge technical aspects of prediction-based systems and ethical debates concerning the use of AI systems: the choice of the socially and normatively significant values that pertain to decision outcomes. If the representation of the decision-making situation were to be very different (e.g., if the decision problem was represented by features that are not quantifiable on a cardinal scale, such as dignity, where we assume that dignity is not quantifiable), we might obtain a different decision outcome. Indeed, as anthropologists and sociologists have argued, the quantitative indicators that represent various phenomena are seductive in their promise to provide concrete knowledge on how the world works. However, they might be unsuccessful in their pairing with context-rich qualitative conceptions of the targets they aim to appropriately and faithfully capture \cite{merry2016seductions}.

\subsection{Training data explanations}

The last piece of explanatory information related to our discussion explores why one or another prediction is generated given what the algorithm has learned from the training data set. During the generation of each data set, practical considerations necessarily play a role in specifying its size and representativeness (e.g., the choice of feature vectors and their target outputs). 

As several have argued, the choice of social and political value-judgments specify what categories the training data set represents, and how it does so \citep{crawford2019excavating}. I call the set of explanatory information that pertain to the specification of the training data set, the training data explanation. If the data set only includes information about the qualification of applicants under the age 24, the algorithm learns different patterns compared to a data set including information about applicants under and above 24. In the case of Nora, for instance, the size and the inclusiveness of training data set play a role in what the algorithm learns, what it recommends, and so provides answers the explanatory question (1).

\section{The many faces of AI explanations}

We have discussed six kinds of explanation of AI outputs, each of which partially explains why a given prediction outcome is reached: three kinds of design explanations capture how design choices determine an AI's decision outcome, and three kinds of explanations provide feature-based information on dependency relations (correlation, causal, and example-based) between particular input and output features. Bringing this together in Figure 4, we obtain a schema that incorporates the six kinds of explanation pertaining to the output generation through an opaque algorithmic system. According to a maximalist conception of explanation, the specific epistemic demands of different stakeholders and the normative significance of the explanations must be factored in. Then, the design and feature-based token information together explains why and how certain algorithmic outcomes are warranted. 

Our framework shows that in different dimensions, design explanations which are not necessarily popular targets of analytical methods are important for the determination of AI outputs. Hence, they should be considered to be part of the philosophical framework for explainable AI systems.

\tikzstyle{my arrow} = [draw=black, very thick, single arrow, minimum height=5.5cm, shape border rotate =#1, fill=gray!10]
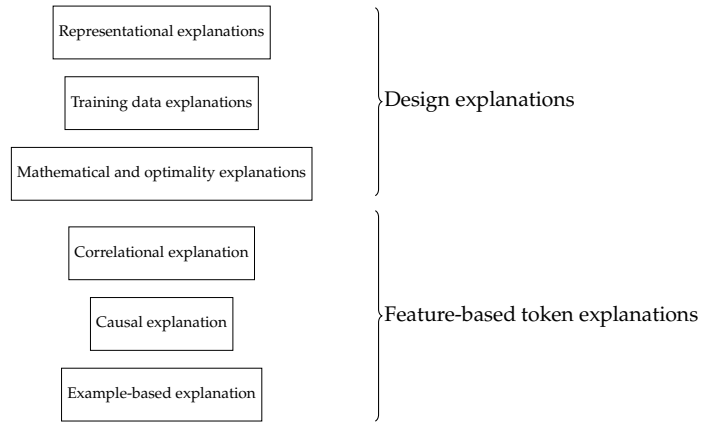
\begin{figure}
\centering
\resizebox{4cm}{!}{\begin{tikzpicture}[every text node part/.style={align=center}]
    \node[main node, rectangle, inner sep=2ex] (0) {\Huge Representational explanations};
    \node[main node, rectangle, inner sep=2ex] (1) [below = 1cm of 0]{\Huge Training data explanations};
    \node[main node, rectangle, inner sep=2ex] (2) [below = 1cm of 1] {\Huge Mathematical and optimality explanations};
    \node[main node, rectangle, inner sep=2ex] (3) [below = 1.5cm of 2]  {\Huge Correlational explanation};
    \node[main node, rectangle, inner sep=2ex] (4) [below = 1cm of 3] {\Huge Causal explanation};
    \node[main node, rectangle, inner sep=2ex] (5) [below = 1cm of 4]  {\Huge Example-based explanation};
\end{tikzpicture}}
\begin{tikzpicture}[overlay]
\path[draw,decorate,decoration={brace,mirror}, right=8pt] (0.75,3) -- (0.75,5.5)
node[midway,right]{\footnotesize Design explanations};
\path[draw,decorate,decoration={brace,mirror}, left=8pt] (0.75,0) -- (0.75,2.8)
node[midway,right]{\footnotesize Feature-based token explanations};
\end{tikzpicture}
\caption{Kinds of AI explanations.}
\end{figure}

Why should we care about this pluralistic picture for AI explanations? Because explanations are reasons, and the explanation for why an ML decision is achieved is determined by what the algorithm has learnt, its reasoning style, and the way that the decision target is represented and measured. Only after these sets of design elements are chosen, can the search for relationships with a data heap begin, and feature-based token explanations can make sense. Hence, design explanations must not be excluded from the exposition of explainable AI. Moreover, the complementary, rather than competitive, nature of explanations distances us from arguing about whether any unique explanation of AI is the right one. 

Why adopt a maximalist conception of AI explanations? The importance of understanding the social and ethical implications of high-stakes AI decision making prevents us from setting minimum requirements on the kinds of explanations an AI system must generate. This insight is deeply rooted in the pragmatist tradition of understanding knowledge, in particular the idea that knowing the world is inseparable from expressed agency within it, and the potential impact of this agency in improving this world in a dynamic way. For example, Dewey \cite{dewey1929quest} draws our attention to the relationship between the legal and economic organization of societies in determining the justification for the production and the use of knowledge for merely economic benefit, rather than the broader benefit of the general public. As this pragmatist approach and the more recent calls for diversifying approaches to the redesign of AI ecosystems (to redistribute power and justice) suggest, an appropriate philosophical framework for intelligent systems that explains their decisions must open sufficient space to show how ethical and social values (as held by different stakeholders) relate to the evaluation of explanations.

Because ML algorithms first learn from data sets that are deliberately curated, the evaluation of the reliability of an algorithmic inference must follow the design choices that are made. The proposed explanatory schema broadens Pearl's ladder of causation \citep{pearl2000causality} by encompassing various design and token feature-based explanations for AI decision outcomes.

While in some cases, the generation of partial explanations is sufficient to satisfy the explanatory goals of a particular end user, a more pluralistic conception of AI explanation would appear to be more useful and trustworthy for assessing the social and ethical implications of AI decision making. This schema, therefore, does not imply that all of these explanations are required in all cases of algorithmic decision making, for all stakeholders, and for all normative and social considerations. When the outcome of an algorithm has no significant impact (such as in classifying images of dogs by breed for entertainment purposes) perhaps the hierarchy of AI explanations does not appear that helpful. However, for consequential decision making in sensitive contexts, where social and ethical challenges are relevant to the opacity of the algorithms, the hierarchy of AI explanations will be very useful. I turn to this topic in the next section.

\section{Background assumptions and value-ladenness of AI inferences}

Identifying the explanations that pertain to AI outcomes does not end the AI explainability debate. The value of such explanations depend on their role in the generation of the understanding of various epistemic, social, and normative considerations. On the one hand, AI explanations must be interpreted and judged according to the epistemic requirements of the stakeholders: operators need to know the description of the inputs that are entered, and the outputs that are generated, whereas decision subjects seek to know what factors contribute to a particular outcome. On the other hand, the relevance and effectiveness of explanations of the consequential decisions made by algorithmic systems should be evaluated in relation to presumed social and ethical values based on which the AI system is designed and the outcomes are determined.

In this section, I develop a philosophical framework for explainable AI that consists of three desiderata: the kinds of explanations discussed in the previous section (reasons), background assumptions that explicitly identify the relevant ethical and social judgments that underlie AI inferences (values), and stakeholders with their epistemic demands. This constitutes the RVS framework for discussing AI explanations. Figure 5 presents a framework that maps each set of values to different types of AI explanations. Stakeholder status can be mapped to kinds of explanations as sketched here, and depending on what matters and how it might matter, stakeholders can challenge assumptions about systems via explanations.

The reasons, as asserted, are the varieties of design and feature-based token explanations. Stakeholders are a broad set of entities that goes beyond those who are directly in charge of system analysis, including a broader group with computer scientists, ethicists, social scientists, policy makers, and end users who seek to make sense of the algorithmic outputs. The use of value judgments entail social and ethical ramifications. I start by giving background to indicate the significance of social and normative values for AI explanations.

Arguments for inherently significant value assumptions in scientific and AI inferences have been thoroughly discussed in the philosophical literature (see, for example, \citep{rudner1953scientist,longino1996cognitive,douglas2000inductive,elliott2017current,levi1960must,Cantwellsmith,dreyfus1992computers}). I start by briefly recalling some of the most prominent of these arguments.

\subsection{Representations: ontological, epistemological, and measurement biases}

As \cite{Cantwellsmith,dreyfus1992computers} point out, several ontological and epistemological assumptions inform the choice of features representative of the decision scenario. Ontological assumptions entail a belief in what there is and what the things are whose existence we take into account. Epistemological assumptions entail that representative features can be sufficiently formalized and quantitatively captured.

For instance, in Nora's case, job candidates can be represented by a set of four features: age, longevity in their previous employment, education level, and gender. Here, we find ontological assumptions: these features exist and are relevant to AI inference making. If we assume that gender (or its proxy) is considered a representative feature, a feature-based token explanation for Nora's rejection is that she is female. Is this a good explanation? No, if we are not legally or ethically motivated to take gender as an existing representative feature for such a decision scenario. This explanation (even if it is true and understandable to the decision subject) is not a good explanation for Nora's rejection. Hence, if the decision subject is aware of the legal and ethical values, she does not accept this explanation even if it is true. Here, the problem is that the AI system makes an incorrect epistemological assumption about the relevance of Nora's gender to its inference making.

In this case, the AI system should not have used this features in the first place. Representational explanations make explicit the set of ontological and epistemological assumptions that are involved in an AI's inference making. The values and assumptions that underpin the choice and utilization of a particular feature in the prediction-based system play a role in the evaluation of a representational explanation of AI.

The way that representational features are measured and quantified is also rooted in epistemological assumptions that have normatively important ramifications. For instance, consider the recent study by \citep{obermeyer2019dissecting} that analyzes healthcare allocation systems to assign scarce and costly healthcare resources to those who would most benefit from them. To resolve this allocation problem with an AI system requires measurement of the benefit of healthcare resources. \citep{obermeyer2019dissecting} showed that when benefits were measured in terms of predictions of healthcare costs, they came with morally significant negative results regarding the decisions for the allocation of resources among Black patients. Therefore, explicitly clarifying how we come to know and make sense of the benefits of health care resources would matter greatly to the evaluation of the system.

Returning to Nora's example, suppose that the algorithm takes four measurable features that describe Nora into account, and finds correlations between the input features and the decision outcome. The correlations constitute the explanation for why a decision outcome is achieved. Let us imagine that a society has strong reason to believe that Nora has the feature ``autonomy to overcome difficult circumstances,'' that this feature matters for hiring purposes, that Nora's background proves she has used her autonomy to overcome extremely difficult circumstances, and that this autonomy is not derivable from the quantified input features required by the decision-making algorithm. Thus, society then cannot accept the explanation that Nora is rejected due to a mismatch between its background assumptions (autonomy as a non-quantifiable feature relevant to the decision-making context) and the four salient, quantified input features used in decision making.

The way that significant features of a population are defined and sampled is rooted in ontological and epistemological assumptions such as what the relevant features of the population are, and how we can or should recognize and represent them. Evaluating the value and significance of these explanations matters for a critical analysis of the AI system in a normative and socially important standpoint. These explanations matter not only to policy makers, auditors, and ethicists, but also to responsible system developers who care about the social impacts of their systems. Ultimately, representational explanations make ontological and epistemological assumptions and measurement biases explicit, and enable stakeholders to provide normative reflections on the adequacy of the correspondence between decision target in a sensitive context and the corresponding reasons for why a given AI outcome is achieved.

\subsection{Training data: source, size, inclusiveness, and inductive risk}

There are two broad sources of value judgments that are relevant to the evaluation of training data explanations. 

The first regards practical decisions on the inclusiveness and size of the data, along with the manner in which it is gathered. These practical decisions can have direct or indirect impacts on what the algorithm learns from past data \citep{olteanu2019social}, and how it can draw inductively reliable inferences about unobserved data. For example,  \cite{crawford2019excavating} have provided a detailed analysis of the political and ethical judgments pertaining to the formation of data sets, such as ImageNet for training ML algorithms. Moreover, several sources of social bias may relate to the training of an algorithm and can give these algorithms a biased spin as historical sources of bias seep into data generation and training. Returning to Nora's example, a more culturally and demographically diverse data set could change what the algorithm learns and how it predicts Nora's future success.

The second is rooted in arguments on inductive risk (\citep{rudner1953scientist,douglas2000inductive,steel2013acceptance}), which indicate that scientific inferences are subject to a level of error. The idea is that because no evidence can establish the validity of a prediction with certainty, its acceptance must always carry a degree of risk. Because there are different error types, managing how much and why each error type matters must be subject to ethical, political, and social value judgments more than any epistemic criterion (such as accuracy or consistency). With a stretch of the imagination, this argument applies to cases of algorithmic inference as well, and ties closely to decisions on the sufficiency of training data explanations and feature-based token explanations.

Hence, for instance, for setting up a system to explain the AI prediction leading to Nora's rejection, explanatory responses to these exemplary questions must be incorporated: why does a training data set about others influence the judgment about Nora's application? That is, why does one training data set, rather than another (with having a different size or error distribution) impact the judgment about Nora's application?

\subsection{Mathematical and optimality inferences: No Free Lunch and reasoning style}

There are two main sources of value judgments that are relevant to the evaluation of the acceptability of mathematical and optimality explanations.

The first source is rooted in theoretical results about ML systems known as the No Free Lunch theorems, which indicate the limitations that algorithms have in performing what they can do. For instance, \citep{wolpert1996lack} shows that no supervised learning algorithm universally performs better than any other, and \citep{wolpert1997no} shows that ``if an [optimization] algorithm does particularly well on average for one class of problems then it must do worse on average over the remaining problems.'' This means that it is not possible to choose the algorithm that will be best to solve any problem. Mathematical and optimality explanations reveal how an algorithm's particular setup affects individuals, and whether this setup is the right one in relation to various criteria, such as reliability and robustness.

Second, it is not clear why a mathematical-optimality system should make this type of decision in the first place. Critical scrutiny is required to address questions such as the following: Why does an amalgamation of statistical and optimization-based reasoning should make the decision whether to hire Nora? If the algorithm is justified, why do these and not those hyperparameters create the algorithm? These types of question connect mathematical and optimality explanations to varieties of explicit or implicit endorsed values and value judgments for setting up an algorithmic perspective and why one or another ML outcome is obtained. These value judgments acquire a significant ethical and political weight when ML algorithms are used in sensitive social contexts, such as in parole and bail decisions in the criminal justice system. Determining the basis of ML predictions encodes the normative claim that inferences via this arrangement of mathematical and optimization thinking is the right way to make decisions. Rather than drawing inferences in this way with an ML algorithm, we could of course use a different reasoning style. For instance, we could situate the problem in a unique context and set the standard for a right decision by permitting relevant human hunches and intuitions, cognitive biases, and historical sensitivities to arrive at the decision outcome. As another alternative, we could appeal to a reasoning style that incorporates the two reasoning approaches to decision making, both a statistically optimal one and an intuitive one, to arrive at a decision outcome \citep{meehl1954clinical,kleinmuntz1990clinical,gottfreson2006clinical}. 

The difference between the choices of reasoning style can be clarified by distinguishing between the kinds of discernment capacities that differ between humans and ML systems \citep{Cantwellsmith}. Smith (2019), for instance, reserves the term ``reckoning'' for calculative rationality of present-day algorithms. This capability is empty of any ethical commitment, authenticity, or deep contextual awareness. He uses ``judgement'' to refer to the human capacity for understanding the relations between appearances and reality in an authentic way that grounds one's contextual awareness and ethical commitments. This authentic understanding of the situation of the decision making, which an ML algorithm seemingly lacks, may impact what the decision output is more than a calculative optimal-statistical reasoning style. To make this point more concretely, I give a simple, real-life example of judgment based on a uniquely human capacity. This example is intended to show how non-calculative judgmental capacities, due to their authentic and contextually aware features, might not be taught to a deep neural network, and therefore such an algorithm cannot use them to arrive at its decisions.

The acceptability of statistical-optimality reckoning is tied to the background assumptions on the legitimacy of this style of reasoning as opposed to nonstatistical judgment. A statistical-optimality explanation sheds light on average behavior of a population sample. By definition, statistical information and inferences are about group behaviors. For instance, the law of large numbers and the central limit theorem only hold true for large data corpa, as certain conditions are satisfied by random variables rather than any single individual. Hence, the decision outcomes obtained from statistical inferences could suffer from limitations regarding the explanation of why one decision, at an individualistic level, has been obtained given the behavior of others. More specifically, the hard question remains to be investigated about how a highly probabilistic outcome justifies a decision result at an individual level. Of course, I do not suggest that statistical inferences are problematic in many contexts. Such inferences are relied upon all the time, and indeed, contemporary scientific inquiry is based mainly on this mode of inquiry. Rather, sensitive social decisions might not always demand a scientific mode of reasoning, and making the best decision might require the involvement of several types of expertise to decide whether to adopt statistical reckoning or a non-statistical judgment. To make this point more concrete, I give a simple real-life example of a judgment based on a uniquely human capacity, which probably cannot be taught to a deep neural network.

\tikzstyle{my arrow} = [draw=black, very thick, single arrow, minimum height=5.5cm, shape border rotate =#1, fill=gray!10]
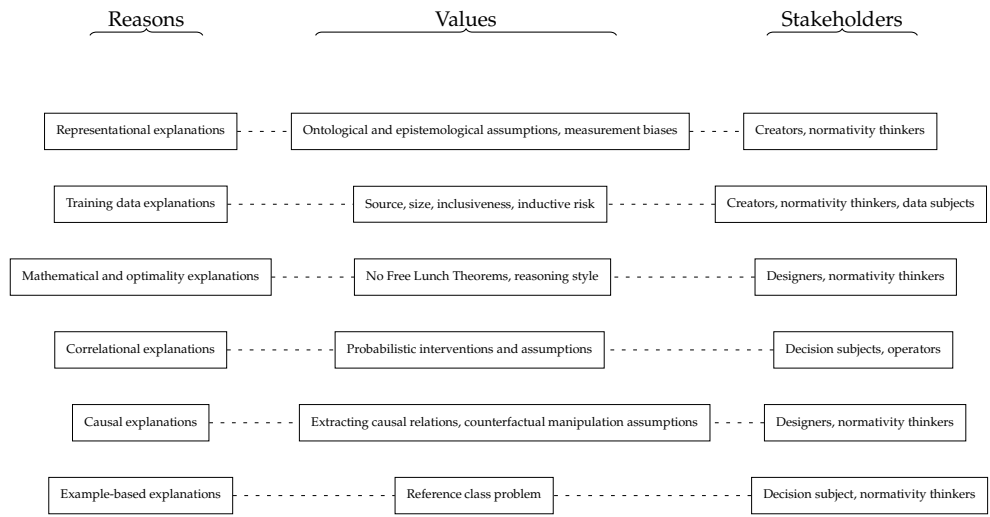
\begin{figure}
\centering
\resizebox{13cm}{!}{\begin{tikzpicture}[every text node part/.style={align=center}]

    \node[main node, rectangle, inner sep=2ex,minimum width=0.5cm,minimum height=1cm] (0) {Representational explanations};
    \node[main node, rectangle, inner sep=2ex,minimum width=0.5cm,minimum height=1cm] (1) [below = 1cm of 0] {Training data explanations};
    \node[main node, rectangle, inner sep=2ex,minimum width=0.5cm,minimum height=1cm] (2) [below = 1cm of 1] {Mathematical and optimality explanations};
    \node[main node, rectangle, inner sep=2ex,minimum width=0.5cm,minimum height=1cm] (3) [below = 1cm of 2]  {Correlational explanations};
    \node[main node, rectangle, inner sep=2ex,minimum width=0.5cm,minimum height=1cm] (4) [below = 1cm of 3] {Causal explanations};
    \node[main node, rectangle, inner sep=2ex,minimum width=0.5cm,minimum height=1cm] (5) [below = 1cm of 4]  {Example-based explanations};

    \node[main node, rectangle, inner sep=2ex,minimum width=0.5cm,minimum height=1cm] (6) [right =3.5cm of 1] {Source, size, inclusiveness, inductive risk};
    \node[main node, rectangle, inner sep=2ex,minimum width=0.5cm,minimum height=1cm] (7) [right =2.3cm of 2] {No Free Lunch Theorems, reasoning style};
    \node[main node, rectangle, inner sep=2ex,minimum width=0.5cm,minimum height=1cm] (8) [right =3cm of 3] {Probabilistic interventions and assumptions};    
    \node[main node, rectangle, inner sep=2ex,minimum width=0.5cm,minimum height=1cm] (9) [right =2.5cm of 4] {Extracting causal relations, counterfactual manipulation assumptions};       
    \node[main node, rectangle, inner sep=2ex,minimum width=0.5cm,minimum height=1cm] (10) [right =4.5cm of 5] {Reference class problem};  
    \node[main node, rectangle, inner sep=2ex,minimum width=0.5cm,minimum height=1cm] (11) [right =1.5cm of 0] {Ontological and epistemological assumptions, measurement biases};
    \path[draw,loosely dashed,thick, right=8pt]
    (1) edge node {} (6);
    \path[draw,loosely dashed,thick, right=8pt]
    (2) edge node {} (7);
    \path[draw,loosely dashed,thick, right=9.5pt]
    (3) edge node {} (8);
    \path[draw,loosely dashed,thick, right=10pt]
    (4) edge node {} (9);
    \path[draw,loosely dashed,thick, right=11pt]
    (5) edge node {} (10);
    \path[draw,loosely dashed,thick, right=11pt]
    (0) edge node {} (11);
    \node[main node, rectangle, inner sep=2ex,minimum width=0.5cm,minimum height=1cm] (12) [right =3cm of 6] {Creators, normativity thinkers, data subjects};
    \node[main node, rectangle, inner sep=2ex,minimum width=0.5cm,minimum height=1cm] (13) [right =4cm of 7] {Designers, normativity thinkers};
    \node[main node, rectangle, inner sep=2ex,minimum width=0.5cm,minimum height=1cm] (14) [right =4.7cm of 8] {Decision subjects, operators};    
    \node[main node, rectangle, inner sep=2ex,minimum width=0.5cm,minimum height=1cm] (15) [right =1.5cm of 9] {Designers, normativity thinkers};       
    \node[main node, rectangle, inner sep=2ex,minimum width=0.5cm,minimum height=1cm] (16) [right =5.5cm of 10] {Decision subject, normativity thinkers}; 
    \node[main node, rectangle, inner sep=2ex,minimum width=0.5cm,minimum height=1cm] (17) [right =1.5cm of 11] {Creators, normativity thinkers};
        \path[draw,loosely dashed,thick, right=8pt]
    (11) edge node {} (17);
    \path[draw,loosely dashed,thick, right=8pt]
    (12) edge node {} (6);
    \path[draw,loosely dashed,thick, right=9.5pt]
    (13) edge node {} (7);
    \path[draw,loosely dashed,thick, right=10pt]
    (14) edge node {} (8);
    \path[draw,loosely dashed,thick, right=11pt]
    (15) edge node {} (9);
    \path[draw,loosely dashed,thick, right=11pt]
    (16) edge node {} (10);
\end{tikzpicture}}
\begin{tikzpicture}[overlay]
\path[draw,decorate,decoration={brace}, above=20pt] (-5,7) -- (-3.5,7)
node[above,midway]{\footnotesize Reasons};
\path[draw,decorate,decoration={brace}, left=8pt] (-2,7) -- (2,7)
node[midway,above]{\footnotesize Values};
\path[draw,decorate,decoration={brace}, left=8pt] (4,7) -- (6,7)
node[midway,above]{\footnotesize Stakeholders};
\end{tikzpicture}
\caption{Reasons, Values, and Stakeholders: The RVS Framework}
\end{figure}

In April 2019, a 96-year-old man, Mr. Coella, was charged with exceeding the speed limit in a school zone in Rhode Island \citep{coella2019}. In court, Mr. Coella explained that he was taking care of his 63-year-old, handicapped son, who has cancer. The reason he gave for exceeding the speed limit is that he was driving his son to the doctor's office. The judge, rather than reaffirming the law to decide Mr. Coella's case, decided that Mr. Coella was ``a good man'' and that his actions were ``what America is all about.'' Accordingly, the judge dismissed the case. It is extremely difficult, if not impossible, to believe that the features captured by the idea of ``goodness of a man'' and ``what America is all about,'' which impacted the decision outcome in this case, could be meaningfully quantified, measured, represented, and learned by a deep neural network. In this example, the judge’s decision output seems to lack any useful appeal of statistical reasoning.

\subsection{Feature-based correlation and causation: cause-effect bias}

Correlational explanations must be evaluated against various probabilistic assumptions regarding the nature of particular random variables, such as, the possible correlation, say, between Nora's ethnicity and her rejection.

The significance of correlational explanations is also tied to a variety of biases that could become insinuated, such as cause-effect bias, the common fallacy that correlation implies causation.

As with causal explanations, some assumptions appear regarding the appropriateness of the use of causal modeling for intervening and manipulating relevant features; for example, it is controversial whether we can really intervene in social categories such as ethnicity or race to generate this type of causation.

\subsection{Assumptions for example-based explanations: reference class problem}

Example-based explanations are judged and evaluated with reference to background assumptions on, for instance, the reference class from which we draw similar instances to the decision outcome. An example-based explanation for why my credit card application is rejected might be that if I had John's age and salary, my application would have been approved. These explanations might be clear to lay people, but they cannot reveal the actual factors on which the decision outcome is based. Keeping the level of discussion at such a level of minimal information might allow private-sector actors to hide other factors that are more like the main reasons for why a certain decision outcome is achieved.

Population selection is important, and the choice of the comparison term for Nora is an instance of the reference class problem: the very choice of what categories to compare to impact the outcome. Several solutions have been posed to the reference class problem; \citep{reichenbach1938probability}, for example, proposes to resolve the problem by picking the narrowest reference class for which there are adequate statistics \citep{Wheeler2016}. How is the narrowest class to be determined? The answer is necessarily loaded with undecidable concerns, such as the criteria of statistical adequacy.

Finally, the RVS framework relates relevant explanations and required value judgments to make sense of them to different stakeholders. Who are the stakeholders? Six main categories are suggested by \cite{tomsett2018interpretable}: decision subjects, data subjects, examiners, creators, operators, and executors. I, however, think that this categorization does not acknowledge the importance of philosophers, social scientists, ethicists, and critical thinkers (a group that I broadly call normativity thinkers), whose aim is to provide a critical analysis of AI systems in relation to values and assumptions. Figure 5 illustrates a set of stakeholders for the assessment of AI explanations and the values judgments required to make sense of the acceptability of explanations of AI inferences. The RVS framework can be used to enable a conversation with the consumers of explanations on reasons and values. The categorization of stakeholders can be adapted to different contexts.

\section{Discussion and conclusion}

Truly opaque AI predictions have the potential to detrimentally affect lives. As a result, explanations for why such predictions are made deserve considerable attention. In this paper, I proposed a multi-dimensional philosophical framework that would bring focused recognition of the importance of the multiplicity of AI explanations, and illustrates how various value judgments, normative considerations, and technical aspects of the ML inferences can be interrelated.

The RVS framework is flexible in accommodating a broader set of technical, legal, and critical stakeholders to evaluate explanations in relating reasons, values, and assumptions required for the acceptability of those reasons as related to different stakeholders. Moreover, the RVS framework for designing intelligent explainable systems facilitates a more dynamic, diverse, and interdisciplinary approach to reflecting, criticizing, mitigating, and improving the algorithmic ecosystem. Because the designers of algorithmic systems, typically, have far more power over the system than those individuals impacted by it, the RVS framework provides a critical landscape for engaging in a more democratic and cooperative way to understand and improve AI ecosystems via AI explanations. This philosophical framework enables and encourages the perspectives of those impacted by an AI system to shape the evaluation of explanations from their system and its decisions. The RVS framework also shows that the set of stakeholders related to algorithmic decisions should become broader and come to incorporate critical and normative disciplines such as philosophy and social and technology studies in their development. The framework is broad and general. It can be shortened and reduced in some contexts if sufficient justification is provided in relation to its shortening.

Where to go next? I believe that the RVS framework can act as a useful schema for designing experiments for human-in-the-loop evaluations of values and acceptability of different AI explanations across a diverse context. For example, we can test whether for a consequential decision context, an evaluation of the quality of AI explanations would change with and without access to the sets of values assumed for the design of the system. Moreover, adapting this framework to the specifics of various decision contexts, such as drone delivery, is an interdisciplinary task to be done.

\bibliography{mybibfile}

\begin{thebibliography}{10}
\expandafter\ifx\csname url\endcsname\relax
  \def\url#1{\texttt{#1}}\fi
\expandafter\ifx\csname urlprefix\endcsname\relax\def\urlprefix{URL }\fi
\expandafter\ifx\csname href\endcsname\relax
  \def\href#1#2{#2} \def\path#1{#1}\fi

\bibitem{cabitza2017unintended}
F.~Cabitza, R.~Rasoini, G.~F. Gensini, Unintended consequences of machine
  learning in medicine, {JAMA} 318~(6) (2017) 517--518.

\bibitem{angwin2016machine}
J.~Angwin, J.~Larson, S.~Mattu, L.~Kirchner, Machine bias: there’s software
  used across the country to predict future criminals, and it’s biased
  against blacks. {P}ropublica 2016 (2016).

\bibitem{bodo2017tackling}
B.~Bod{\'o}, N.~Helberger, K.~Irion, F.~Zuiderveen~Borgesius, J.~Moller,
  B.~van~de Velde, N.~Bol, B.~van Es, C.~de~Vreese, Tackling the algorithmic
  control crisis-the technical, legal, and ethical challenges of research into
  algorithmic agents, Yale Journal of Law \& Technology 19 (2017) 133.

\bibitem{fox2007argumentation}
J.~Fox, D.~Glasspool, D.~Grecu, S.~Modgil, M.~South, V.~Patkar,
  Argumentation-based inference and decision making--a medical perspective,
  IEEE intelligent systems 22~(6) (2007) 34--41.

\bibitem{ribeiro2016should}
M.~T. Ribeiro, S.~Singh, C.~Guestrin, Why should i trust you?: Explaining the
  predictions of any classifier, in: Proceedings of the 22nd ACM SIGKDD
  international conference on knowledge discovery and data mining, ACM, 2016,
  pp. 1135--1144.

\bibitem{veale2018clarity}
M.~Veale, L.~Edwards, Clarity, surprises, and further questions in the article
  29 working party draft guidance on automated decision-making and profiling,
  Computer Law \& Security Review 34~(2) (2018) 398--404.

\bibitem{gunning2017explainable}
D.~Gunning, Explainable artificial intelligence (xai), Defense Advanced
  Research Projects Agency (DARPA), nd Web 2 (2017) 2.

\bibitem{adadi2018peeking}
A.~Adadi, M.~Berrada, Peeking inside the black-box: A survey on explainable
  artificial intelligence ({XAI}), IEEE Access 6 (2018) 52138--52160.

\bibitem{swartout1993explanation}
W.~R. Swartout, J.~D. Moore, Explanation in second generation expert systems,
  in: Second generation expert systems, Springer, 1993, pp. 543--585.

\bibitem{binns2018s}
R.~Binns, M.~Van~Kleek, M.~Veale, U.~Lyngs, J.~Zhao, N.~Shadbolt, 'it's
  reducing a human being to a percentage' perceptions of justice in algorithmic
  decisions, in: Proceedings of the 2018 CHI Conference on Human Factors in
  Computing Systems, 2018, pp. 1--14.

\bibitem{chen2014situation}
J.~Y. Chen, K.~Procci, M.~Boyce, J.~Wright, A.~Garcia, M.~Barnes, Situation
  awareness-based agent transparency, Tech. rep., Army research lab aberdeen
  proving ground md human research and engineering (2014).

\bibitem{chouldechova2017fairer}
A.~Chouldechova, M.~G'Sell, Fairer and more accurate, but for whom?, arXiv
  preprint arXiv:1707.00046.

\bibitem{hayes2017improving}
B.~Hayes, J.~A. Shah, Improving robot controller transparency through
  autonomous policy explanation, in: 2017 12th ACM/IEEE International
  Conference on Human-Robot Interaction, IEEE, 2017, pp. 303--312.

\bibitem{kemper2019transparent}
J.~Kemper, D.~Kolkman, Transparent to whom? no algorithmic accountability
  without a critical audience, Information, Communication \& Society 22~(14)
  (2019) 2081--2096.

\bibitem{kim2015interactive}
B.~Kim, Interactive and interpretable machine learning models for human machine
  collaboration, Ph.D. thesis, Massachusetts Institute of Technology (2015).

\bibitem{lepri2018fair}
B.~Lepri, N.~Oliver, E.~Letouz{\'e}, A.~Pentland, P.~Vinck, Fair, transparent,
  and accountable algorithmic decision-making processes, Philosophy \&
  Technology 31~(4) (2018) 611--627.

\bibitem{mercado2016intelligent}
J.~E. Mercado, M.~A. Rupp, J.~Y. Chen, M.~J. Barnes, D.~Barber, K.~Procci,
  Intelligent agent transparency in human--agent teaming for multi-uxv
  management, Human factors 58~(3) (2016) 401--415.

\bibitem{hoffman2018explaining}
R.~R. Hoffman, G.~Klein, S.~T. Mueller, Explaining explanation for
  ``explainable ai'', in: Proceedings of the Human Factors and Ergonomics
  Society Annual Meeting, Vol.~62, SAGE Publications Sage CA: Los Angeles, CA,
  2018, pp. 197--201.

\bibitem{zhang2020effect}
Y.~Zhang, Q.~V. Liao, R.~K. Bellamy, Effect of confidence and explanation on
  accuracy and trust calibration in ai-assisted decision making, in:
  Proceedings of the Conference on Fairness, Accountability, and Transparency,
  2020, p. 295–305.

\bibitem{lipton2016mythos}
Z.~C. Lipton, The mythos of model interpretability, arXiv preprint
  arXiv:1606.03490.

\bibitem{doshi2017towards}
F.~Doshi-Velez, B.~Kim, Towards a rigorous science of interpretable machine
  learning, arXiv preprint arXiv:1702.08608.

\bibitem{guidotti2018survey}
R.~Guidotti, A.~Monreale, S.~Ruggieri, F.~Turini, F.~Giannotti, D.~Pedreschi, A
  survey of methods for explaining black box models, ACM Computing Surveys
  (CSUR) 51~(5) (2018) 1--42.

\bibitem{miller2019explanation}
T.~Miller, Explanation in artificial intelligence: Insights from the social
  sciences, Artificial Intelligence 267 (2019) 1--38.

\bibitem{zerilli2019transparency}
J.~Zerilli, A.~Knott, J.~Maclaurin, C.~Gavaghan, Transparency in algorithmic
  and human decision-making: is there a double standard?, Philosophy \&
  Technology 32~(4) (2019) 661--683.

\bibitem{zednik2019solving}
C.~Zednik, Solving the black box problem: A normative framework for explainable
  artificial intelligence, Philosophy \& Technology (2019) 1--24.

\bibitem{tomsett2018interpretable}
R.~Tomsett, D.~Braines, D.~Harborne, A.~Preece, S.~Chakraborty, Interpretable
  to whom? a role-based model for analyzing interpretable machine learning
  systems, arXiv preprint arXiv:1806.07552.

\bibitem{sep-causation-counterfactual}
P.~Menzies, H.~Beebee, {Counterfactual Theories of Causation}, in: E.~N. Zalta
  (Ed.), The {Stanford} Encyclopedia of Philosophy, summer 2020 Edition,
  Metaphysics Research Lab, Stanford University, 2020.

\bibitem{erhan2009visualizing}
D.~Erhan, Y.~Bengio, A.~Courville, P.~Vincent, Visualizing higher-layer
  features of a deep network, University of Montreal 1341~(3) (2009) 1.

\bibitem{berk2013statistical}
R.~A. Berk, J.~Bleich, Statistical procedures for forecasting criminal
  behavior: A comparative assessment, Criminology \& Pub. Pol'y 12 (2013) 513.

\bibitem{goldstein2015peeking}
A.~Goldstein, A.~Kapelner, J.~Bleich, E.~Pitkin, Peeking inside the black box:
  Visualizing statistical learning with plots of individual conditional
  expectation, Journal of Computational and Graphical Statistics 24~(1) (2015)
  44--65.

\bibitem{letham2015interpretable}
B.~Letham, C.~Rudin, T.~H. McCormick, D.~Madigan, et~al., Interpretable
  classifiers using rules and bayesian analysis: Building a better stroke
  prediction model, The Annals of Applied Statistics 9~(3) (2015) 1350--1371.

\bibitem{ribeiro2016model}
M.~T. Ribeiro, S.~Singh, C.~Guestrin, Model-agnostic interpretability of
  machine learning, arXiv preprint arXiv:1606.05386.

\bibitem{sundararajan2017axiomatic}
M.~Sundararajan, A.~Taly, Q.~Yan, Axiomatic attribution for deep networks, in:
  Proceedings of the 34th International Conference on Machine Learning-Volume
  70, 2017, pp. 3319--3328.

\bibitem{dabkowski2017real}
P.~Dabkowski, Y.~Gal, Real time image saliency for black box classifiers, in:
  Advances in Neural Information Processing Systems, 2017, pp. 6967--6976.

\bibitem{lundberg2017unified}
S.~M. Lundberg, S.-I. Lee, A unified approach to interpreting model
  predictions, in: Advances in neural information processing systems, 2017, pp.
  4765--4774.

\bibitem{kim2018}
B.~Kim, M.~Wattenberg, J.~Gilmer, C.~Cai, J.~Wexler, F.~Viegas, R.~Sayres,
  Interpretability beyond feature attribution: Quantitative testing with
  concept activation vectors (tcav), in: International Conference on Machine
  Learning, 2018, pp. 2673--2682.

\bibitem{hara2016making}
S.~Hara, K.~Hayashi, Making tree ensembles interpretable, arXiv preprint
  arXiv:1606.05390.

\bibitem{wachter2017counterfactual}
S.~Wachter, B.~Mittelstadt, C.~Russell, Counterfactual explanations without
  opening the black box: Automated decisions and the gpdr, Harvard Journal of
  Law \& Technology 31 (2017) 841.

\bibitem{lakkaraju2017learning}
H.~Lakkaraju, C.~Rudin, Learning cost-effective and interpretable treatment
  regimes, in: Artificial Intelligence and Statistics, 2017, pp. 166--175.

\bibitem{zhao2019causal}
Q.~Zhao, T.~Hastie, Causal interpretations of black-box models, Journal of
  Business \& Economic Statistics (2019) 1--19.

\bibitem{kim2014bayesian}
B.~Kim, C.~Rudin, J.~A. Shah, The bayesian case model: A generative approach
  for case-based reasoning and prototype classification, in: Advances in Neural
  Information Processing Systems, 2014, pp. 1952--1960.

\bibitem{kim2016examples}
B.~Kim, R.~Khanna, O.~O. Koyejo, Examples are not enough, learn to criticize!
  criticism for interpretability, in: Advances in Neural Information Processing
  Systems, 2016, pp. 2280--2288.

\bibitem{li2018deep}
O.~Li, H.~Liu, C.~Chen, C.~Rudin, Deep learning for case-based reasoning
  through prototypes: A neural network that explains its predictions, in:
  Thirty-Second AAAI Conference on Artificial Intelligence, 2018, pp.
  3530--3537.

\bibitem{krizhevsky2012imagenet}
A.~Krizhevsky, I.~Sutskever, G.~E. Hinton, Imagenet classification with deep
  convolutional neural networks, in: Advances in neural information processing
  systems, 2012, pp. 1097--1105.

\bibitem{sabour2017dynamic}
S.~Sabour, N.~Frosst, G.~E. Hinton, Dynamic routing between capsules, in:
  Advances in neural information processing systems, 2017, pp. 3856--3866.

\bibitem{kosiorek2019stacked}
A.~Kosiorek, S.~Sabour, Y.~W. Teh, G.~E. Hinton, Stacked capsule autoencoders,
  in: Advances in Neural Information Processing Systems, 2019, pp.
  15486--15496.

\bibitem{smith1994law}
M.~L. Smith, S.~A. Kane, The law of large numbers and the strength of
  insurance, in: Insurance, risk management, and public policy, Springer, 1994,
  pp. 1--27.

\bibitem{lecun2015deep}
Y.~LeCun, Y.~Bengio, G.~Hinton, Deep learning, Nature 521~(7553) (2015) 436.

\bibitem{hempel1948studies}
C.~G. Hempel, P.~Oppenheim, Studies in the logic of explanation, Philosophy of
  Science 15~(2) (1948) 135--175.

\bibitem{hempel1965aspects}
C.~G. Hempel, Aspects of Scientific Explanation; And Other Essays in the
  Philosophy of Science, New York, Free Press, 1965.

\bibitem{mittelstadt2019explaining}
B.~Mittelstadt, C.~Russell, S.~Wachter, Explaining explanations in ai, in:
  Proceedings of the conference on fairness, accountability, and transparency,
  ACM, 2019, pp. 279--288.

\bibitem{salmon1984scientific}
W.~C. Salmon, Scientific explanation and the causal structure of the world,
  Princeton University Press, 1984.

\bibitem{strevens2008depth}
M.~Strevens, Depth: An account of scientific explanation, Harvard University
  Press, 2008.

\bibitem{potochnik2007optimality}
A.~Potochnik, Optimality modeling and explanatory generality, Philosophy of
  Science 74~(5) (2007) 680--691.

\bibitem{bokulich2011scientific}
A.~Bokulich, How scientific models can explain, Synthese 180~(1) (2011) 33--45.

\bibitem{batterman2014minimal}
R.~W. Batterman, C.~C. Rice, Minimal model explanations, Philosophy of Science
  81~(3) (2014) 349--376.

\bibitem{chirimuuta2017explanation}
M.~Chirimuuta, Explanation in computational neuroscience: Causal and
  non-causal, The British Journal for the Philosophy of Science 69~(3) (2017)
  849--880.

\bibitem{lange2016because}
M.~Lange, Because without cause: Non-causal explanations in science and
  mathematics, Oxford University Press, 2016.

\bibitem{pincock2007role}
C.~Pincock, A role for mathematics in the physical sciences, No{\^u}s 41~(2)
  (2007) 253--275.

\bibitem{reutlinger2018explanation}
A.~Reutlinger, J.~Saatsi, Explanation beyond causation: philosophical
  perspectives on non-causal explanations, Oxford University Press, 2018.

\bibitem{rice2015moving}
C.~Rice, Moving beyond causes: Optimality models and scientific explanation,
  No{\^u}s 49~(3) (2015) 589--615.

\bibitem{batterman2001devil}
R.~W. Batterman, The devil in the details: Asymptotic reasoning in explanation,
  reduction, and emergence, Oxford University Press, 2001.

\bibitem{lange2013really}
M.~Lange, Really statistical explanations and genetic drift, Philosophy of
  Science 80~(2) (2013) 169--188.

\bibitem{johnston2019boeing}
P.~Johnston, R.~Harris, The boeing 737 max saga: lessons for software
  organizations, Software Quality Professional 21~(3) (2019) 4--12.

\bibitem{mau2019metric}
S.~Mau, The metric society: On the quantification of the social, John Wiley \&
  Sons, 2019.

\bibitem{merry2016seductions}
S.~E. Merry, The seductions of quantification: Measuring human rights, gender
  violence, and sex trafficking, University of Chicago Press, 2016.

\bibitem{crawford2019excavating}
K.~Crawford, T.~Paglen, Excavating {AI}: The politics of images in machine
  learning training sets, \emph{Excavating AI}.

\bibitem{dewey1929quest}
J.~Dewey, The quest for certainty, 3rd Edition, New York: Capricorn Books,
  1960.

\bibitem{pearl2000causality}
J.~Pearl, Causality: models, reasoning and inference, Vol.~29, Springer, 2000.

\bibitem{rudner1953scientist}
R.~Rudner, The scientist qua scientist makes value judgments, Philosophy of
  science 20~(1) (1953) 1--6.

\bibitem{longino1996cognitive}
H.~E. Longino, Cognitive and non-cognitive values in science: Rethinking the
  dichotomy, in: Feminism, science, and the philosophy of science, Springer,
  1996, pp. 39--58.

\bibitem{douglas2000inductive}
H.~Douglas, Inductive risk and values in science, Philosophy of science 67~(4)
  (2000) 559--579.

\bibitem{elliott2017current}
K.~C. Elliott, D.~Steel, Current controversies in values and science, Taylor \&
  Francis, 2017.

\bibitem{levi1960must}
I.~Levi, Must the scientist make value judgments?, The Journal of Philosophy
  57~(11) (1960) 345--357.

\bibitem{Cantwellsmith}
B.~C. Smith, The promise of Artificial Intelligence, MIT University Press,
  2019.

\bibitem{dreyfus1992computers}
H.~L. Dreyfus, L.~Hubert, et~al., What computers still can't do: A critique of
  artificial reason, MIT press, 1992.

\bibitem{obermeyer2019dissecting}
Z.~Obermeyer, B.~Powers, C.~Vogeli, S.~Mullainathan, Dissecting racial bias in
  an algorithm used to manage the health of populations, Science 366~(6464)
  (2019) 447--453.

\bibitem{olteanu2019social}
A.~Olteanu, C.~Castillo, F.~Diaz, E.~Kiciman, Social data: Biases,
  methodological pitfalls, and ethical boundaries, Frontiers in Big Data 2
  (2019) 13.

\bibitem{steel2013acceptance}
D.~Steel, Acceptance, values, and inductive risk, Philosophy of Science 80~(5)
  (2013) 818--828.

\bibitem{wolpert1996lack}
D.~H. Wolpert, The lack of a priori distinctions between learning algorithms,
  Neural computation 8~(7) (1996) 1341--1390.

\bibitem{wolpert1997no}
D.~H. Wolpert, W.~G. Macready, No free lunch theorems for optimization, IEEE
  transactions on evolutionary computation 1~(1) (1997) 67--82.

\bibitem{meehl1954clinical}
P.~E. Meehl, Clinical versus statistical prediction: A theoretical analysis and
  a review of the evidence., University of Minnesota Press, 1954.

\bibitem{kleinmuntz1990clinical}
B.~Kleinmuntz, Clinical and actuarial judgment, Science 247~(4939) (1990)
  146--146.

\bibitem{gottfreson2006clinical}
S.~D. Gottfreson, L.~J. Moriarty, Clinical versus actuarial judgments in
  criminal justice decisions: Should one replace the other, Fed. Probation 70
  (2006) 15.

\bibitem{coella2019}
L.~Moran, Judge tosses speeding ticket of 96-year-old man caring for son with
  cancer., Hoffington Post.

\bibitem{reichenbach1938probability}
H.~Reichenbach, On probability and induction, Philosophy of Science 5~(1)
  (1938) 21--45.

\bibitem{Wheeler2016}
G.~Wheeler, Machine epistemology and big data, in: L.~McIntyre, A.~Rosenberg
  (Eds.), The Routledge Companion to The Philosophy of Social Science,
  Routledge: London, UK, 2016, pp. 321--329.

\end{thebibliography}

\end{document}